\documentclass{osa-article}

\journal{oe}

\articletype{Research Article}
\usepackage{lineno}

\begin{document}

\title{Compact, Portable, Thermal-Noise-Limited Optical Cavity with Low Acceleration Sensitivity}

\author{Megan Lauree Kelleher,\authormark{1,2,$\dagger$} Charles A.McLemore,\authormark{1,2}, Dahyeon Lee, \authormark{1,2} Josue Davila-Rodriguez, \authormark{2,4} Scott A. Diddams, \authormark{1,2,3}, and Franklyn Quinlan \authormark{2,*}}

\address{\authormark{1}Department of Physics, 390 UCB, University of Colorado, Boulder, CO 80309, USA\\
\authormark{2}National Institute of Standards and Technology, 325 Broadway, Boulder, CO 80305, USA\\
\authormark{3}Electrical, Computer and Energy Engineering, 425 UCB, University of Colorado, Boulder, CO 80309, USA\\
\authormark{4}current address: Stable Laser Systems, Boulder, CO 80301, USA}
\email{\authormark{$\dagger$}megan.kelleher@colorado.edu \\ 
\authormark{*}franklyn.quinlan@nist.gov}



\begin{abstract*}
We develop and demonstrate a compact (less than $6$~mL) portable Fabry-P\'{e}rot optical reference cavity.  A laser locked to the cavity is thermal noise limited at $2\times10^{-14}$ fractional frequency stability.  Broadband feedback control with an electro-optic modulator enables near thermal-noise-limited phase noise performance from $1$~Hz to $10$~kHz offset frequencies. The additional low vibration, temperature, and holding force sensitivity of our design makes it well suited for out-of-the-lab applications such as optically derived low noise microwave generation, compact and mobile optical atomic clocks, and environmental sensing through deployed fiber networks.
\end{abstract*}

\section{Introduction}
Lasers frequency-locked to the resonance of a vacuum-gap Fabry-P\'{e}rot (FP) optical cavity have reached fractional frequency instability lower than $1\times10^{-16}$ at $1$~s of averaging time \cite{Hafner:15, PhysRevLett.118.263202}. The frequency instability is inherited from the fluctuations in cavity length as sampled by an optical beam. The sampled length fluctuations can be reduced to less than the diameter of a proton. Such extraordinary performance in laser frequency stability has aided in the advancement of state-of-the-art laboratory optical atomic clocks, with applications in the redefinition of the SI second \cite{Riehle_2018} and tests of fundamental physics \cite{Ratio2021}. The best fractional frequency stabilities in cavity-stabilized lasers are achieved by exploiting long cavity lengths (up to $\sim48$~cm \cite{lvarez_2019,Hafner:15}), operation at cryogenic temperatures \cite{PhysRevLett.118.263202}, and extensive environmental isolation by way of vibration-insensitive mounting and multiple layers of thermal isolation \cite{Sanjuan:19, doi:10.1063/1.4941718}. However, many out-of-the-lab applications of stable lasers, such as portable optical atomic clocks\cite{Ludlow15}, earthquake detection using undersea optical fiber\cite{Marra18}, and low phase noise microwave generation via optical frequency division (OFD) \cite{Fortier2011, Xie2017}, benefit from the high stability available in the optical domain, but are incompatible with the size, weight, and infrastructure requirements of large or cryogenic cavity systems. Furthermore, short cavities have the potential for inherently low acceleration sensitivity without the need for active vibration stabilization\cite{PhysRevA.74.053801}. 

The search for laser frequency reference cavities that are rigidly held, have reduced size and weight, and can operate in harsh and unpredictable environments has led to the development of both solid-state dielectric resonators \cite{lee_suh_chen_li_diddams_vahala_2013, zhang2019microrod, stern2020ultra, loh2020operation, alnis2011thermal}, and compact vacuum-gap FPs \cite{Davila-Rodriguez:17, Leibrandt:11, Webster:11, Didier:18, Argence:12,Sanjuan:19, Herbers:22}. Solid-state dielectric resonators are impressively small, typically millimeter-scale, and, in some cases, can be manufactured at scale. However, these resonators suffer from higher thermorefractive noise and temperature sensitivity that has limited the fractional frequency instability to the $10^{-13}$ level and above \cite{alnis2011thermal}. By placing the optical mode in vacuum and using low expansion materials, compact and rigidly held FPs can reach fractional frequency stabilities $\sim1\times10^{-15}$ with a cavity volume near $60$~mL \cite{Davila-Rodriguez:17}. Notably, simulations of the noise of vacuum-gap FPs predict that their size can be reduced to only a few milliliters while maintaining fractional frequency stability performance well below the $10^{-13}$ level of the best solid-state dielectric resonators; indeed, a recent demonstration of a 10 mm-long, $8$ mL-volume cavity reached $6\times10^{-15}$ \cite{PhysRevApplied.18.054054}, albeit without testing of holding force or acceleration sensitivity. Thus, there remains a compelling performance space that can be achieved with a compact FP optical frequency reference, provided that a laser locked to the FP can operate with noise at or near the cavity thermal noise limit, and the FP has low acceleration and holding force sensitivity.

Here we present a simple, rigidly held cylindrical vacuum-gap FP cavity reaching a fractional frequency stability of $2\times10^{-14}$, capable of supporting applications in low phase noise microwave generation via optical frequency division (OFD)\cite{Fortier2011, Xie2017}, distributed optical fiber sensing \cite{doi:10.1063/1.5113955,9.1117/12.2059448,1440515}, and mobile optical clocks\cite{Ludlow15}. The cavity is composed of fused silica (FS) mirrors and an ultralow expansion (ULE) glass spacer which is only 6.3 mm long. The cavity volume is $5.2$~mL. A laser locked to the cavity operates at the cavity thermal noise limit for noise offset frequencies ranging from $\sim1$~Hz to $\sim10$~kHz. To our knowledge, the phase noise level at 10~kHz, at approximately $-108$~dBc/Hz on the optical carrier, is one of the lowest reported for any vacuum-gap FP \cite{Bondu:96, Davila-Rodriguez:17}, or dielectric resonator \cite{Liu:22,Liang2015}. If paired with an optical frequency comb, the laser system can support state-of-the-art microwave phase noise that is comparable to the lowest phase noise achieved to date for offset frequencies above $\sim100$~Hz. Measurements of the cavity's low sensitivity to holding force indicate the cavity may be reliably held on its end faces. Though other groups have simulated and studied its impact \cite{fasano_2021,wiens_schiller_2018, Leibrandt:11, Webster:11}, these are the first direct measurements of holding force sensitivity of which we are aware. Additionally, the cavity acceleration sensitivity for three mechanical axes was measured to be $5 \times 10^{-11}g^{-1}$, $3 \times 10^{-10}g^{-1}$, and $6 \times 10^{-10}g^{-1}$. A variation on our design also allowed us to explore trade-offs in holding force sensitivity, noise, and long-term stability for a cavity composed only of ULE and larger diameter mirrors. In the following section, we discuss the cavity design and laser locking approach. In Section 3, we present cavity holding force and acceleration sensitivity simulations and measurement results, and measurements on the phase noise and frequency stability of a laser locked to the cavity. In Section 4 we summarize and conclude. 

\section{Cavity Design \& Laser Locking}

\begin{figure}
\centering\includegraphics[width=12cm]{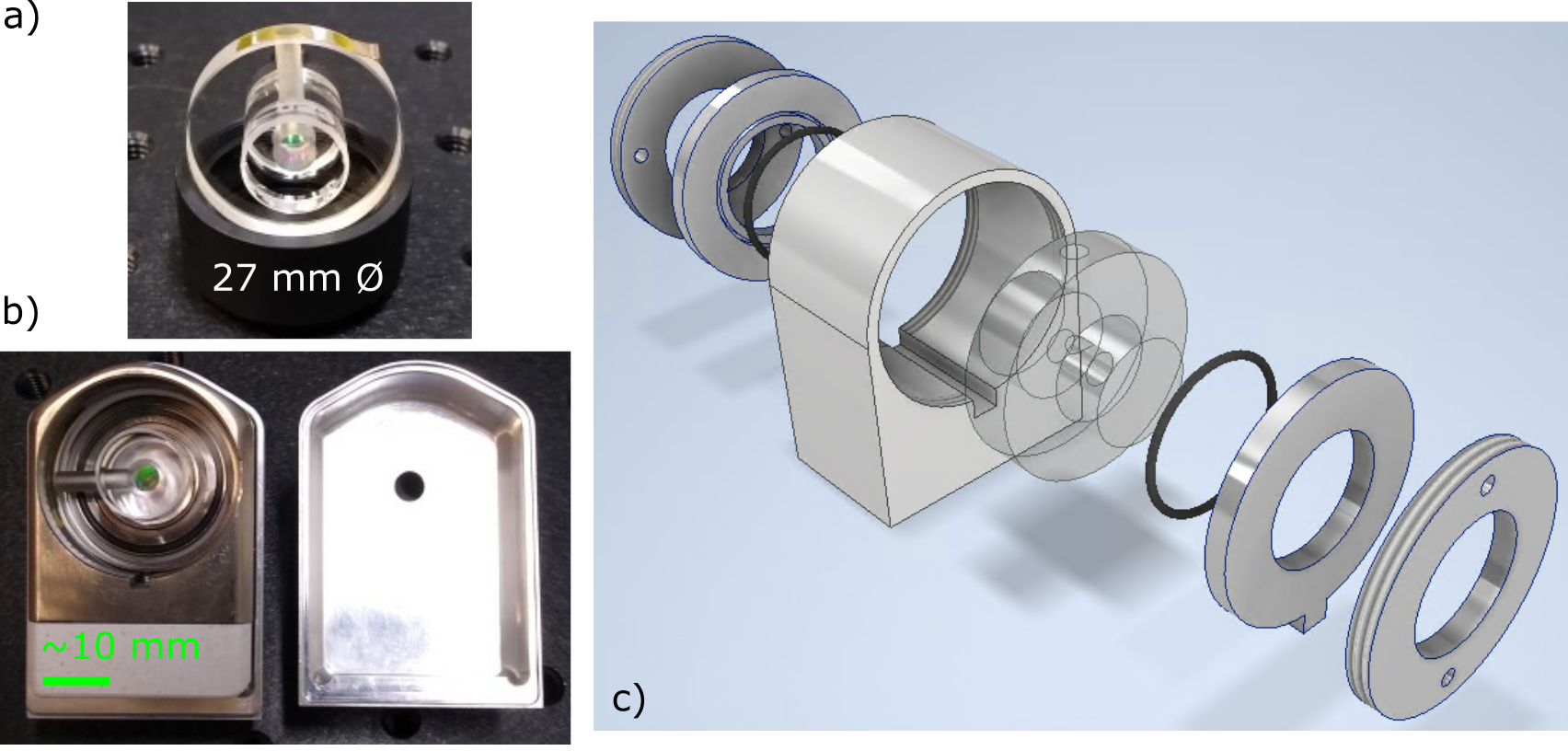}
\caption{a) Photograph of the optical cavity, which is $27$~mm in diameter. The spacer is 6.3 mm thick and the mirrors are $12.7$~mm in diameter and $6.35$~mm thick. b)~Photograph of the optical cavity in its mounting structure. The mounting structure sits on top of a Macor spacer and is placed inside of a heat shield, also pictured. c)~Exploded cartoon view of the holding structure. The cavity is suspended in the Invar mounting structure by two Viton o-rings. These o-rings are held in place by two Invar holders. Note the notch on the holder that prevents the o-ring from rotating when the structure is assembled. Finally, threaded retaining rings are applied to rigidly hold the cavity in place. }
\label{fig:photos}
\end{figure}

Compact and portable FP reference cavity designs must balance noise performance, temperature sensitivity, acceleration sensitivity, tolerance to mirror misalignment \cite{Leibrandt:11}, sensitivity to changes in the holding force, and manufacturability. We have chosen a simple cylindrical geometry for ease of manufacture and its high degree of mechanical symmetry (see Fig. \ref{fig:photos}). Mechanical symmetry is important for maintaining low acceleration sensitivity, and has been exploited in many compact cavity geometries, including cubes \cite{Webster:11}, spheres \cite{Leibrandt:11}, pyramids \cite{Didier:18}, and other compact cylinders \cite{Davila-Rodriguez:17}. The ULE spacer material is chosen for its low thermal expansion and measures $27$~mm in diameter and $6.3$~mm in length. The spacer has a single vent hole with a diameter of $3$~mm for evacuating the cavity. The mirror substrates are FS with a diameter of $12.7$~mm and a standard thickness of $6.35$~mm. Both mirrors have a radius of curvature (ROC) of $1$~m with a flat outer contact annulus for optical contact bonding to the spacer. The total cavity volume is only $5.2$~mL. The mirrors were manufactured and polished in the same batch, leading to nearly equal optical contact areas. As we show below, the asymmetry introduced by the single vent hole increases the acceleration sensitivity of the cavity. 

The mirror material was chosen by considering the thermal noise and temperature sensitivity. Fused silica has a higher mechanical quality factor than ULE, and results in lower thermal noise \cite{PhysRevLett.93.250602}. For the lowest temperature sensitivity, the cavity should be operated at its zero-crossing temperature ($T_{zc}$), where the linear coefficient of thermal expansion (CTE) passes through zero \cite{Fox:09}. The $T_{zc}$ of ULE is typically near room temperature. However, employing FS mirrors shifts the $T_{zc}$ of the cavity to well below room temperature due to distortions of the mirror shape caused by the comparatively large radial expansion of the mirrors \cite{Legero:10}. This effect is particularly important for short FP cavities, where the mirror distortions are a larger fraction of the cavity length. To counteract this effect, ULE backing rings \cite{Legero:10} can be used to shift the cavity $T_{zc}$ back to a convenient temperature, but it is not always possible to fully compensate the large $T_{zc}$ shift of compact cavities in this way. High $T_{zc}$ ULE \cite{PhysRevApplied.18.054054} can also be used but is not typically specified to the required precision for a repeatable compact cavity manufacturing process. Here, using a ULE spacer with unknown $T_{zc}$, we designed our system to rely on temperature stabilization and shielding of the cavity enclosure. This was driven by our targeted application of low noise microwave generation and undersea fiber optic cable measurements, where short-term phase and frequency fluctuations are of greater concern than long-term cavity length drift.

The $1$~m ROC of the mirrors was chosen to maximize the optical spot size on the mirrors, $w$, while maintaining reasonable tolerance to mirror misalignment. For an optical cavity whose thermal noise is dominated by Brownian noise in the coatings (as is the case here), the phase noise power scales as $1/w^2$ \cite{PhysRevD.42.2437}, and for cavities where the ROC is much larger than the cavity length (L), $w^2$ is proportional to $\sqrt{\text{ROC}}$. However, the larger ROC results in a larger optical axis displacement due to a mirror tilt or displacement, and the shift in the optical axis away from the cavity mechanical axis increases the effective acceleration sensitivity of the cavity \cite{nazarova_riehle_sterr_2006}. Again assuming $\text{ROC} \gg L$, the beam displacement on the mirror surfaces $d$ is given by $d\approx 0.5\times \text{ROC}\times\theta$, where $\theta$ is the mirror tilt angle \cite{waldman_2009}. The residual angle between the faces of our spacer is $< 50~\mu$rad, leading to a maximum beam displacement near $25~\mu$m. Furthermore, sag in the mirror surface for large ROC mirrors is extremely small when the mirror diameter is small. For example, the sag at the center of a $12.7$~mm diameter mirror with $1$~m ROC is only $\sim20~\mu$m. This makes creating a contact annulus with the required roughness and surface figure without spoiling the smoothness of the center of the mirror extremely difficult.  Indeed, the largest ROC we could obtain on $12.7$~mm diameter mirrors with an annulus for optical contact bonding was 1 m. 

The cavity’s rigid holding structure was designed to minimize the effect of holding force variation on cavity length, shown in Fig. \ref{fig:photos}. We used finite element analysis (FEA) software to design a holding geometry that can provide first-order insensitivity to the holding force. Given the short length of our cavity spacer, the cavity is held on the spacer end faces with Viton o-rings. The o-rings are held against the cavity with a backing plate, behind which is a threaded piece that screws into a holding mount. The backing plates and holding mount are made of low expansion Invar to minimize temperature changes of the holding structure from coupling to changes to the cavity holding force. The base of the holding mount is composed of Macor to reduce the thermal conductivity from the outer vacuum enclosure. Additionally, a polished aluminum heat shield is placed around the cavity and holding structure to reduce the radiative heat transfer to the cavity. The volume enclosed by the heat shield is $\sim$40 mL.  

\begin{figure}
\centering\includegraphics[width=13cm]{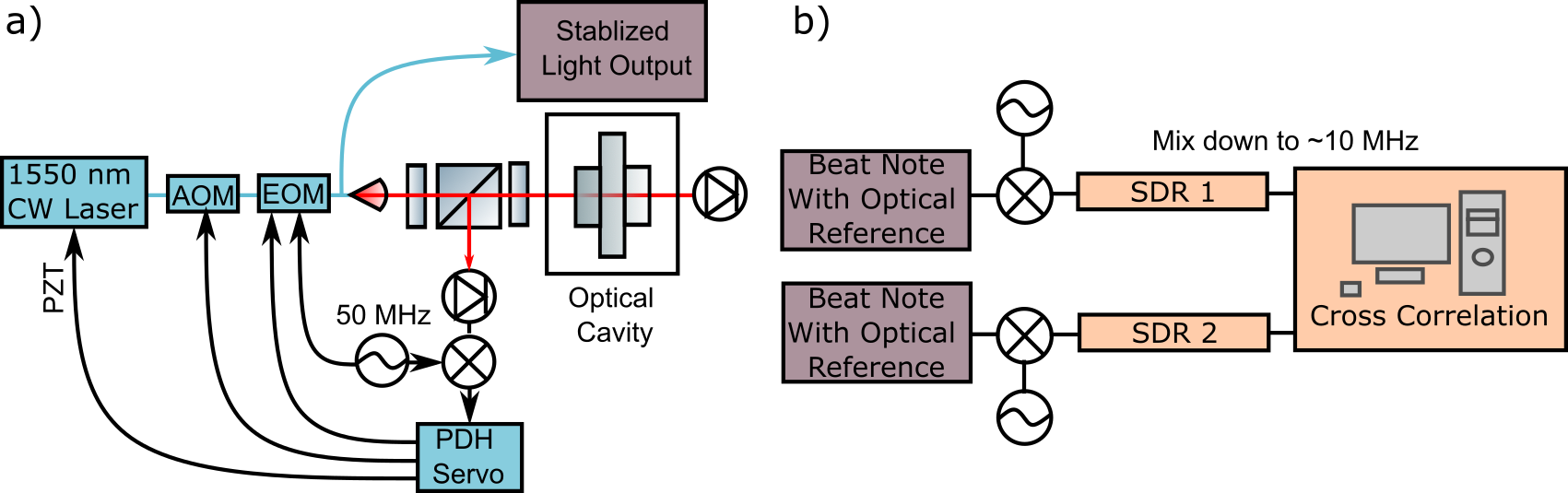}
\caption{a) Simplified block diagram of the optical cavity system. A $1550$~nm laser is sent through a fiberized AOM then EOM. The light is then split in fiber. Some of the stabilized light is sent to the optical cavity. The $1550$~nm laser is stabilized to the optical cavity resonance using feedback to the laser PZT, AOM, and EOM. A frequency diplexer (not shown) is used at the rf input of the EOM so that the EOM can be used for both the PDH sideband generation (at $50$~MHz) and feedback control. The transmission detector can provide a RIN servo error signal, but is only utilized during flip tests to stabilize the intracavity power because the alignment is affected by the optical breadboard flexing. b) Stabilized laser light is sent on to be compared to two different optical references. One is another cavity-stabilized laser at 1550~nm. The other is a comb that is frequency-referenced to a cavity-stabilized 1156~nm laser \cite{Schioppo2017}. Software defined radio (SDR) is used to track the phase of the signals, which are then cross-correlated to remove uncommon noise in the frequency combs and beat note detection.}
\label{fig:SYSdiagram}
\end{figure}
\section{Results}
 
The laser frequency was locked to the cavity with the Pound-Drever-Hall (PDH) technique \cite{Black:2001}, as shown in Fig. \ref{fig:SYSdiagram}. Cavity ring-down measurements yielded a finesse of 600,000, providing a steep discriminator slope for laser locking.  We employed a broadband locking scheme utilizing feedback to an electro-optic modulator (EOM) to achieve nearly cavity thermal noise-limited performance over a broad offset frequency range \cite{Endo:18, Hall:84, Bondu:96}. Light from a 1550~nm commercial fiber laser is routed through a fiber-coupled acousto-optic modulator (AOM) frequency shifter, followed by a fiber-coupled EOM and a 90/10 fiber coupler, then is launched into free space to interrogate the cavity. A free-space circulator directs light reflected from the cavity to a photodetector used for PDH frequency stabilization. A separate photodetector placed at the back end of the cavity is used to stabilize the intracavity power, though during normal operation it is not needed as it was determined that our laser's intensity noise does not contribute to the measured phase noise. Laser frequency stabilization to the cavity is implemented through three feedback paths: a piezoelectric transducer controlling the laser cavity length is used for low bandwidth/large dynamic range frequency corrections, the AOM for mid-bandwidth (to few 100~kHz) feedback, and the EOM is used for high bandwidth (up to $1$~MHz) feedback. The same EOM is also used to impart 50~MHz phase modulation sidebands on the laser light for the PDH error signal generation. The optical power impinging on the cavity is $\sim600~\mu$W, leaving greater than $10$~mW of frequency-stabilized output from the 90/10~coupler. Since we use a single EOM for both PDH phase modulation and feedback control, the stabilized light output will have 50~MHz sidebands. This will be well outside the bandwidth of any frequency comb lock, and is not anticipated to impede any applications using a comb. If stabilized light without PDH sidebands is desired, PDH phase modulation and feedback control can be separated by using two EOMs. 

To measure the laser phase noise, the stabilized laser output was split and heterodyned against two separate ultrastable optical frequency references. The heterodyne beat notes were simultaneously digitally sampled by software defined radio (SDR), and the phase fluctuations were extracted \cite{Sherman2016}. The phase noise cross-spectrum was then obtained by averaging the complex product of the Fourier transforms of the individual phase records \cite{Davila-Rodriguez:17}. This allowed us to reject the noise of the optical phase references and proved to be particularly important to reveal the low phase noise of our system for offset frequencies $>1$~kHz.

\subsection{Holding Force and Acceleration Sensitivity}

\begin{figure}
\centering\includegraphics[width=12cm]{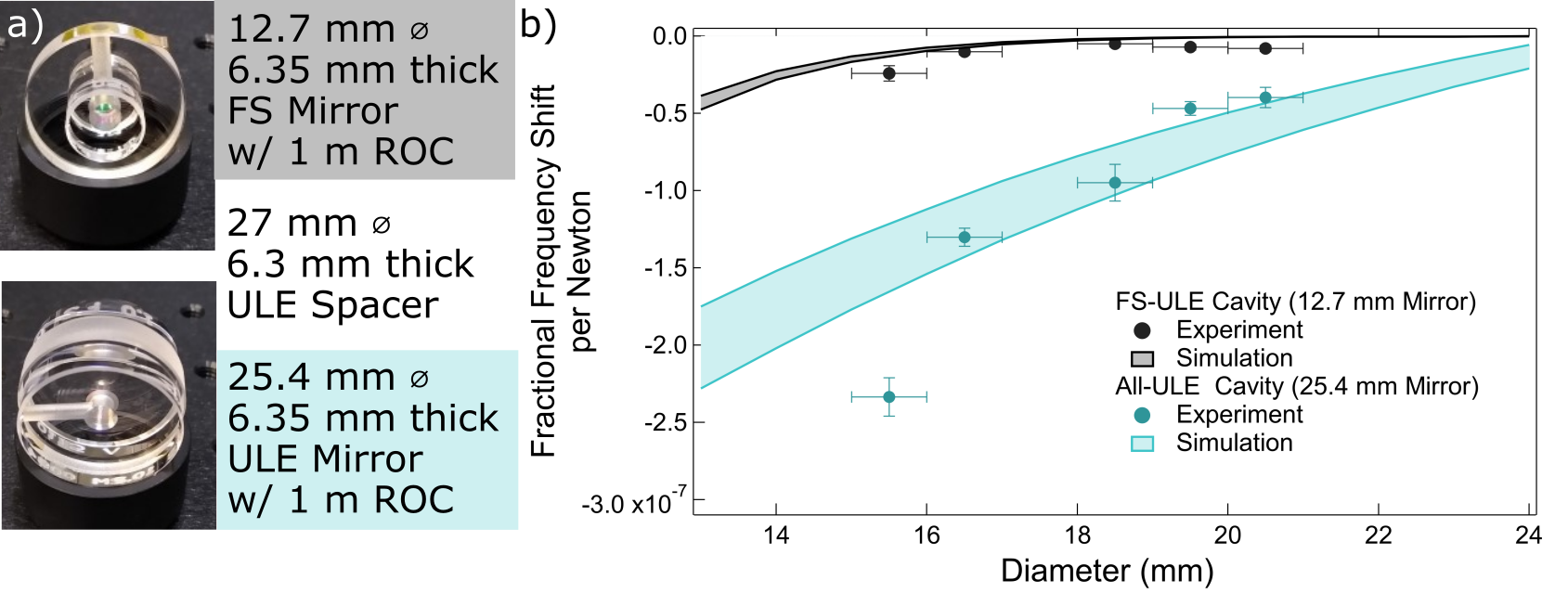}
\caption{a) Photographs of the primary cavity and the supplemental all-ULE cavity. The cavities have identical spacers, but different mirrors. b)~Experimental results showing the holding force sensitivity’s dependence on o-ring size, demonstrating excellent agreement with the finite element analysis results, which are shown as shaded regions. The y-error bars are a combination of the standard deviation in the frequency shift measurement and the uncertainty in the mass of the weight applied. This does not include any systematic offsets. The x-axis uncertainty is an estimate based on the thickness of the o-ring (1 mm). Simulations incorporated a range of mirror-spacer contact areas, where the contact annulus width varied from 3.2 mm to 2.5 mm in the FS-ULE cavity, and 6.4 mm to 5.1 mm in the all-ULE cavity. }
\label{fig:holdingforce}
\end{figure}

Sensitivity to holding force was measured and compared to simulations, as shown in Fig. \ref{fig:holdingforce}. The cavity is held by rings on the end faces of the spacer, and can be first-order insensitive to changes to the holding force. This can be understood by considering the behavior of a linear elastic cylinder compressed on the end faces either near the center or the outer diameter. For a small ring-radius holding force, the length of cylinder’s central axis will reduce as the holding force is increased. In contrast, when squeezed at a large diameter near the cylinder’s rim, distortions in the spacer shape cause the length along the central axis to increase. A first-order holding force-insensitive point on cavity length lies in between these two extremes. This simple description is complicated by the fact that spacer distortions are coupled to mirror distortions, such that the mirror diameter, thickness and the contact area between the mirror and spacer impact the holding force sensitivity. Moreover, depending on the spacer’s aspect ratio, the holding force insensitive radius may be too close to the outer rim to be practicable.
 
To simulate this holding force sensitivity for our design, an axisymmetric cavity model was built in FEA simulation software, and a simulated force was applied to the ULE spacer end faces along a $\sim$0.2 mm-wide ring that is equal and opposite on both ends of the cavity. Cavity length changes were calculated as a function of the holding force ring radius. As expected, the holding force sensitivity depends on many parameters such as mirror thickness and diameter, spacer thickness and diameter, and contact area. For an optical contact annulus width of 6.4 mm, a holding-force zero-crossing is predicted for a holding diameter just above 20.5 mm, slightly larger than our largest o-ring.  Perhaps more importantly, simulations of this cavity geometry show a weak dependence of the holding force sensitivity to the o-ring diameter. This can benefit manufacturability because it can be difficult to obtain an o-ring with diameter that exactly matches that of the zero-crossing. To further verify our holding force results, we simulated and measured the holding force sensitivity of an all-ULE cavity consisting of 25.4 mm diameter mirrors with 1~m ROC, also shown in Fig. \ref{fig:holdingforce}. In this case, the holding rings are placed on the backside of the mirrors. For both cavity designs, the dependence of holding force sensitivity on the contact area is displayed using the shaded sections. For each simulation, the upper bound of the shaded region represents a larger contact area, and the lower bound represents a smaller contact area. The primary cavity design (FS-ULE) has a lesser dependence on contact area and is more resilient to manufacturing error in this way. 

To measure the holding force sensitivity, the cavity was mounted in air with the optical axis aligned vertically (parallel to the force due to gravity). A small weight of known mass was applied to the top of the cavity for each available holding diameter. The change in frequency of a laser locked to the cavity was observed, and the resulting fractional frequency shift per newton was calculated. The weight was applied and removed a minimum of eight times to establish statistics on the reproducibility of holding force sensitivity measurement. Potential systematic errors, such as a small angle between the optical axis and the force due to gravity, are ignored. 
 
The results are consistent with the simulations, verifying our understanding of the cavity and its design. The primary cavity design with $12.7$~mm FS mirrors has a fractional frequency shift per newton of less than $2.4\times 10^{-8}$ with all of the o-ring diameters that were tested. This behavior is contrasted with the secondary cavity with $25.4$~mm ULE mirrors, where the holding force sensitivity has a stronger dependence on o-ring diameter. These simulations also agree with the holding force sensitivity measurements. 

We assess the holding force sensitivity by considering a change in holding force due to a temperature change of the rigid mounting structure. By accounting for the CTE of the Invar structure and the elastic properties of the o-rings, we estimate the temperature-dependent holding force change on the cavity as $\sim3$~mN/~${}^{\circ}$C. This leads to a fractional cavity length change of $< 10^{-10}/{}^{\circ}$C. This is several orders of magnitude smaller than the CTE of the FS-ULE cavity of $\sim10^{-7}/{}^{\circ}$C. This is also smaller than what one could expect when operating near the cavity $T_{zc}$, where the CTE for a 1${}^{\circ}$C change is $\sim10^{-9}/{}^{\circ}$C. Thus, the temperature induced cavity length change will be dominated by the cavity itself as opposed to the holding force changes.

\begin{figure}
\centering\includegraphics[width=12cm]{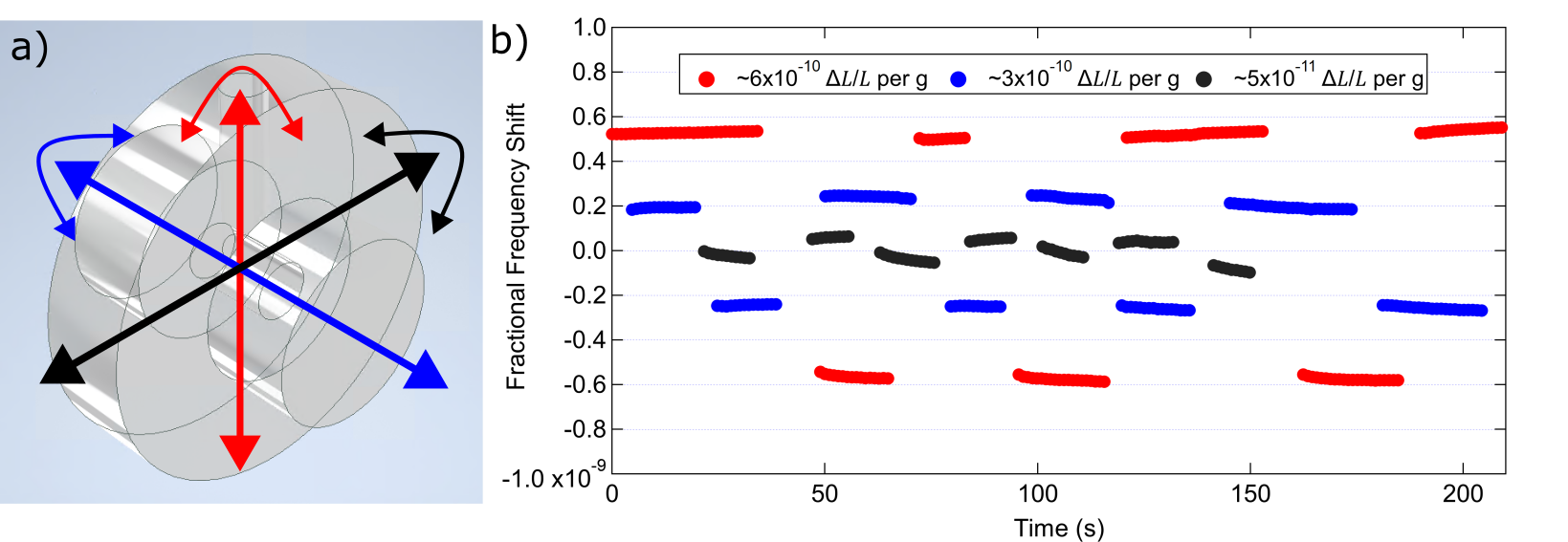}
\caption{a) Diagram of the cavity indicating the mechanical axis along which the cavity is flipped. In the red flip test, the cavity vent hole goes from up to down. In the blue flip test, the optical axis is flipped, and in the black flip test, the mechanical axis which is perpendicular to both the vent hole and the optical beam is flipped. b)~Flip test measurements of the primary cavity acceleration sensitivity. The cavity is flipped along three mechanical axes inducing a 2g acceleration change while a laser stays locked to the cavity. The change in frequency is observed. Linear drift, attributed to cavity temperature change, is removed from the data. The worst axis is $6\times10^{-10} \frac{\Delta L}{L}$ per g, and this axis corresponds to flipping the vent hole up and down. The asymmetry caused by the vent hole is likely contributing to higher acceleration sensitivity along this axis.}
\label{fig:flip}
\end{figure}

Previous work on cavities ranging from 5 cm to 20 cm in length with cubic, spherical, and cylindrical geometries have demonstrated passive (without vibration feedforward correction) acceleration sensitivities in the $10^{-11}g^{-1}$ to mid $10^{-10}g^{-1}$ range \cite{Herbers:22,PhysRevA.87.023829,Webster:11,Li:18,Argence:12,Chen:14, Davila-Rodriguez:17}. We simulated and measured the acceleration sensitivity of our FS-ULE cavity under a 2g static flip test \cite{Webster:11,Davila-Rodriguez:17, PhysRevA.87.023829}. We placed the cavity in vacuum onto a rotatable optical breadboard that allowed us to flip the cavity along three mechanical axes. The results of this measurement are in Fig. \ref{fig:flip}. The measured acceleration sensitivities along the three axes are $5 \times 10^{-11}g^{-1}$, $3 \times 10^{-10}g^{-1}$, and $6 \times 10^{-10}g^{-1}$. The axis with the highest sensitivity corresponds to a flip that changes the cavity vent hole from pointing up to pointing down. This is the largest mechanical asymmetry in the design, and FEA simulations indicate that the acceleration sensitivity should be at best $\sim3 \times 10^{-10}g^{-1}$ along this mechanical axis. Furthermore, we note that this low level of acceleration sensitivity was achieved without active alignment of the cavity supports. The cavity is simply centered in the mount by visual inspection and the Viton o-rings are centered to the mount using groves in the holding structure. Simulations indicate that additional minor asymmetries would result in higher acceleration sensitivity. For example, offsets in the mirrors along the vent hole direction can cause $\sim6 \times 10^{-10}g^{-1}$ per millimeter offset,  and offsets in the holding rings  along the vent hole direction result in $\sim8 \times 10^{-11}g^{-1}$ per millimeter offset. A similar geometry with smaller vent holes evenly spaced radially should perform better due to a higher degree of symmetry. In the high symmetry cavity case, alignment tolerance will dominate, likely at the $10^{-11}g^{-1}$ level assuming sub-millimeter alignment tolerance, but this depends on a wide parameter space and could likely be further optimized by changing variables like mirror thickness.  

\subsection{Phase Noise and Fractional Frequency Stability} 

\begin{figure}
\centering\includegraphics[width=12cm]{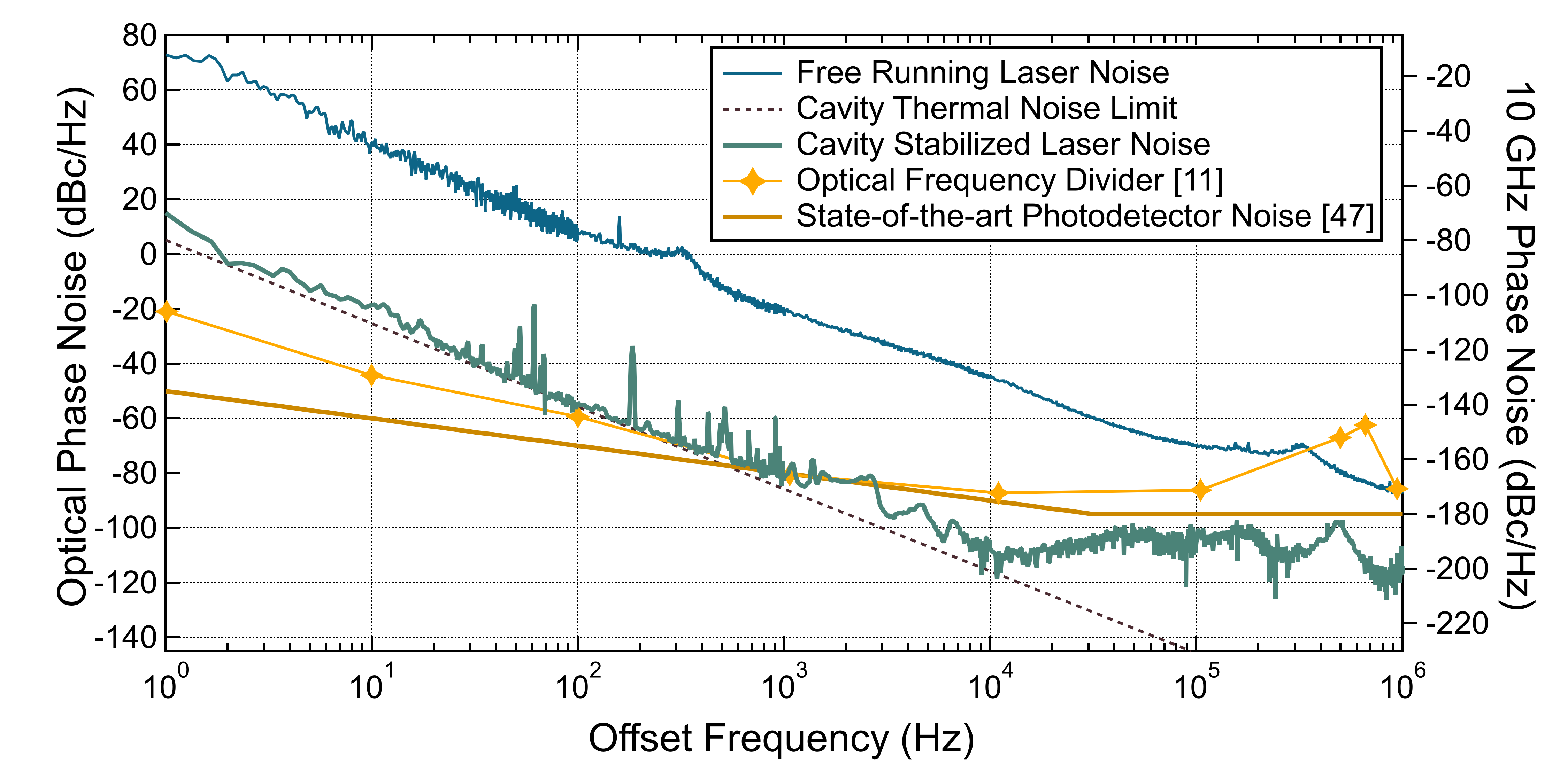}
\caption{Measured phase noise of the free-running commercial laser in blue, cavity-stabilized laser phase noise in green, the cavity thermal noise limit in black, an optical frequency divider in gold \cite{Xie2017}, and state-of-the-art photodetector phase noise at 10~GHz in brown \cite{Baynes:15}. Note that the noise of the cavity-stabilized laser stays near the thermal noise limit to $\sim10$~kHz offset. Because the phase noise of the cavity-locked laser above 1~kHz offset frequency is at or below a state-of-the-art photodetector level, this cavity can serve as a compact, low-noise reference for OFD microwave generation.}
\label{fig:PN}
\end{figure}

Phase noise of the cavity-stabilized laser, the predicted cavity thermal noise limit, and the laser’s free-running noise, are shown in Fig. \ref{fig:PN}. The integrated timing jitter from 1~MHz to 1.3~Hz is 12~fs. The laser phase noise closely follows the cavity thermal noise for offset frequencies between 1~Hz and 10~kHz. The predicted thermal noise is calculated using FEA software \cite{Kessler:12} and includes Brownian noise of the spacer, mirror coatings, and mirror substrates, as well as thermoelastic noise of the substrates and spacer. Only the Brownian noise of our small spacer differed significantly from simple analytic models. Thermo-optic noise was calculated analytically \cite{BRAGINSKY2000303, PhysRevLett.91.260602, Cole2013}, but did not make a substantial contribution to the total thermal noise. Broadband noise reduction of the free-running laser noise was realized using feedback to the EOM, resulting in a gain bandwidth $>$ 1 MHz. At 10 kHz offset, this large feedback bandwidth provides $>$ 60 dB suppression of the free-running laser noise. To our knowledge, the measured phase noise level at 10 kHz is one of the lowest reported for any vacuum-gap FP (despite the cavity's small size) \cite{Bondu:96, Davila-Rodriguez:17} or dielectric resonator \cite{Liu:22,Liang2015} .

When coupled with an optical frequency comb, this cavity can support low-noise microwave generation through OFD \cite{Fortier2011}. The phase noise contribution of our cavity on a 10 GHz carrier is shown on the right axis of Fig. \ref{fig:PN}. For comparison, the phase noise of a state-of-the-art 10 GHz photodetector \cite{Baynes:15}, and one of the lowest noise OFD systems demonstrated to date are also shown \cite{Xie2017}. Importantly, our compact cavity can support microwaves comparable to that of the lowest noise OFD signals produced yet-to-date for offset frequencies greater than $\sim$100~Hz. Above $\sim$1 kHz, the microwave noise contribution from the cavity is below that of the projected photodetector noise, and thus will not adversely impact the signal. Above 10~kHz offset frequencies, the cavity supports 10~GHz phase noise below -180~dBc/Hz. Of course, the residual noise of the optical frequency comb will also contribute to the final microwave phase noise; however, these results show that an extremely compact cavity can enable microwave signals whose noise is competitive with those systems that are referenced to much larger-size optical cavity systems.

\begin{figure}
\centering\includegraphics[width=12cm]{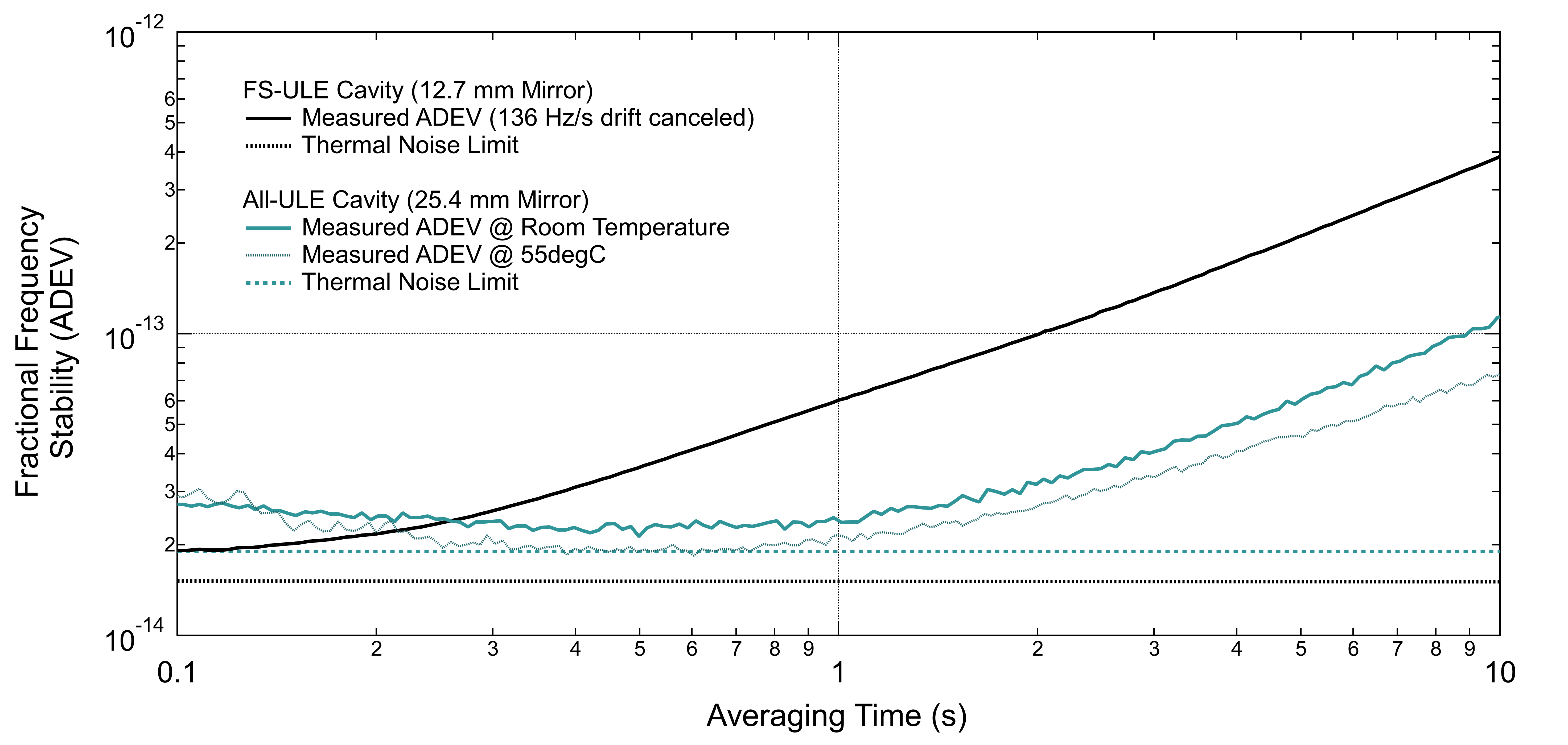}
\caption{Measured fractional frequency stability expressed as an Allan deviation (ADEV) of a laser locked to the primary cavity and the all-ULE cavity. An external AOM was used to compensate for a linear drift of 136~Hz/s on the locked laser. There was a residual drift of 14 Hz/s at the time of the measurement. The all-ULE cavity did not have a drift canceling AOM and the drift rate was substantially lower, 3~Hz/s at room temperature and 2 Hz/s at 55~${}^{\circ}$C. Note that the all-ULE Cavity is thermal noise limited at 1~s at 55~${}^{\circ}$C. }
\label{fig:ADEV}
\end{figure}

Figure \ref{fig:ADEV} shows the measured fractional frequency stability of our primary (FS mirrors and ULE spacer) cavity, given in terms of the Allan deviation (ADEV). For comparison, the ADEV of the all-ULE cavity is also shown. The ADEV is calculated using the phase record of the SDR measuring the beat note of our cavities against a comb that is frequency-referenced to a cavity-stabilized 1156~nm laser. For the primary cavity, an external AOM driven by a direct digital synthesizer (DDS) with a linearly chirped frequency correction was used to compensate for the 136~Hz/s linear drift, which is likely due to the CTE of this cavity. A residual drift of 14~Hz/s remained after compensation, limited by the frequency resolution of the DDS. The fractional frequency stability of the primary cavity (black curve) is near its calculated thermal noise limit (dashed black curve) at 0.1~s averaging time. The slightly larger thermal noise limit of the all-ULE cavity is shown in dashed blue. The ADEV of the all-ULE cavity was measured at both room temperature near 23~${}^{\circ}$C and at 55~${}^{\circ}$C, demonstrating the large range of temperatures at which this cavity can operate (our setup did not allow for cooling the cavity below room temperature). At room temperature, the frequency is nearly thermal noise limited from 0.1~s to 1~s with a linear drift of 3~Hz/s. At 55~${}^{\circ}$C, the frequency stability reaches the calculated thermal noise limit from 0.3~s to 0.8~s, and exhibits a slightly lower drift rate of 2~Hz/s. No $T_{zc}$ was found with this cavity below 55~${}^{\circ}$C, though we note the ULE used for this spacer is legacy material from previous experiments \cite{Fox:09}, and its material properties are not well known. Still, despite the higher thermal noise and lack of $T_{zc}$, the all-ULE design can be a valuable compromise where long term-stability and low drift are desirable.

\section{Conclusions}
We have demonstrated a compact optical frequency reference cavity design, with a cavity volume of $5.2$~mL, compatible with out-of-the-lab applications which can support ultralow noise microwave generation through OFD. The design uses a simple cylindrical geometry for ease of manufacturing and is rigidly held. Using a frequency-locking technique utilizing EOM feedback with $>$1~MHz bandwidth, we demonstrated near thermal noise-limited optical phase noise performance, reaching -110~dBc/Hz at 10~kHz offset. Despite the higher thermal noise floor resulting from its compact size, this system provides one of the lowest phase noise results at 10 kHz offset for any optical reference cavity of which we are aware. For larger offsets, the noise remains below -100~dBc/Hz. Moreover, the cavity demonstrated near thermal noise limited fractional frequency stability of $2\times10^{-14}$ at 0.1~s. A lower frequency drift rate was achieved with an all-ULE cavity variation of the design.

Additionally, a low holding force sensitivity of the cavity was measured, demonstrating agreement with simulations. Importantly, the holding force sensitivity showed minimal dependence on the holding radius. Measurements of the acceleration sensitivity ranged from  $\sim6\times10^{-10}g^{-1}$ for the cavity’s mechanical axis that displays the largest asymmetry to $\sim5 \times10^{-11}g^{-1}$ for the least sensitive mechanical axis. This low acceleration sensitivity was achieved with a simple holding geometry and minimal alignment of the cavity into the mount.

 Further improvements to the cavity performance are straightforward. The acceleration sensitivity can be reduced by implementing a more symmetric vent hole pattern in the cavity spacer. An all-ULE version of the 12.7 mm cavity design could combine the low dependence of the holding force diameter with low frequency drift (at a minimal cost to the low phase noise). With its demonstrated low noise performance, our cavity design fills an important gap in the performance-size trade space, enabling compact out-of-the-lab systems with improved phase and frequency stability.

\begin{backmatter}

\bmsection{Funding} This work is funded by AFRL and NIST. 

\bmsection{Acknowledgments} We thank R. Fox for providing the ULE used in these experiments, and the NIST Yb optical atomic clock group for providing frequency-stabilized light at 1156~nm. We also thank Andrew Ludlow, Tobias Bothwell, and Laura Sinclair for their insightful comments on the manuscript.

This work is a contribution of an agency of the U.S. government and is not subject to copyright in the USA. Commercial equipment is identified for scientific clarity only and does not represent an endorsement by NIST. 

\bmsection{Disclosures} The authors declare no conflicts of interest.

\bmsection{Data availability} Data underlying the results presented in this study are not publicly available at this time but may be obtained from the authors upon a reasonable request.

\end{backmatter}

\bibliography{sample}

\end{document}